\documentclass[aps,pra,twocolumn,superscriptaddress,floatfix,amsmath,amssymb]{revtex4-2}
\usepackage{amsbsy}
\usepackage{amstext}
\usepackage{braket}
\usepackage[unicode=true,pdfusetitle,
 bookmarks=false,
 breaklinks=false,pdfborder={0 0 1},backref=false,colorlinks=false]
 {hyperref}
 \DeclareMathOperator{\sech}{sech}

\hypersetup{colorlinks,citecolor=NavyBlue,filecolor=black,linkcolor=black,urlcolor=black}
\usepackage[dvipsnames]{xcolor}

\hypersetup{colorlinks=true,citecolor=NavyBlue,filecolor=black,linkcolor=black,urlcolor=black}

\makeatletter

\@ifundefined{textcolor}{}
{%
 \definecolor{BLACK}{gray}{0}
 \definecolor{WHITE}{gray}{1}
 \definecolor{RED}{rgb}{1,0,0}
 \definecolor{GREEN}{rgb}{0,1,0}
 \definecolor{BLUE}{rgb}{0,0,1}
 \definecolor{CYAN}{cmyk}{1,0,0,0}
 \definecolor{MAGENTA}{cmyk}{0,1,0,0}
 \definecolor{YELLOW}{cmyk}{0,0,1,0}
}

\usepackage{amsmath}
\usepackage{graphicx}
\usepackage{amssymb}
\usepackage{txfonts,color}
\usepackage{dsfont}
\makeatother

\begin{document}


\title{Multiplexed quantum state transfer in waveguides}

\author{Guillermo F. Pe{\~n}as}\email{guillermof.pens.fdez@iff.csic.es}
\affiliation{Instituto de F{\'i}sica Fundamental, IFF-CSIC, Calle Serrano 113b, 28006 Madrid, Spain}
\author{Ricardo Puebla}
\affiliation{Departamento de F{\'i}sica, Universidad Carlos III de Madrid, Avda. de la Universidad 30, 28911 Legan{\'e}s, Spain}
\author{Juan Jos\'e Garc\'ia-Ripoll}
\affiliation{Instituto de F{\'i}sica Fundamental, IFF-CSIC, Calle Serrano 113b, 28006 Madrid, Spain}

\begin{abstract}
In this article, we consider a realistic waveguide implementation of a quantum network that serves as a testbed to show how to maximize the storage and manipulation of quantum information in QED setups. We analyze two approaches using wavepacket engineering and quantum state transfer protocols. First, we propose and design a family of orthogonal photons in the time domain. These photons allow for a selective interaction with distinct targeted qubits. Yet, mode multiplexing employing resonant nodes is largely spoiled by cross-talk effects. This motivates the second approach, namely, frequency multiplexing. Here we explore the limits of frequency multiplexing through the waveguide, analyzing its capabilities to host and faithfully transmit photons of different frequencies within a given bandwidth. We perform detailed one- and two-photon simulations and provide theoretical bounds for the fidelity of coherent quantum state transfer protocols under realistic conditions. Our results show that state-of-the-art experiments can employ dozens of multiplexed photons with global fidelities fulfilling the requirements imposed by fault-tolerant quantum computing. This is with the caveat that the conditions for single-photon fidelity are met.
\end{abstract}

\maketitle

\section{Introduction}
Superconducting circuit and waveguide quantum electrodynamics (QED)~\cite{Blais_cQED_review,Sheremet_WQED_review} address two promising and synergistic platforms in quantum science and engineering. While circuit QED is primarily concerned with quantum information processing in superconducting chips, and issues such as gate fidelity and scalability, waveguide QED setups enable the distribution of quantum resources among those processors, via two-way transduction between stationary qubits and propagating photons~\cite{Pechal2013, Leung2019, Magnard2020, Kannan2020}, entanglement sharing~\cite{Narla2016, Kurpiers2017, Zhong2021, Qiu2023, Storz2023}, and the engineering of large quantum many-body states~\cite{Schwartz2016, Ferreira2022}. The synergistic combination of both setups opens a realistic path towards the implementation of distributed quantum computing~\cite{Cirac1999, Beals2013, Cacciapuoti2020, Caleffi2022} and the achievement of large-scale quantum information processing architectures~\cite{Ladd2010, Dowling2003, Wehner2018}.

Within this context, in which we engineer and model quantum links between quantum hubs as waveguide QED setups~\cite{Eichler2012, Pechal2013, Zeytinoglu2015, Kurpiers2017, Bienfait2019, Magnard2020, Storz2023, Penas2022, Penas2023, Qiu2023, Grebel2024}, enlarging the information capacity of the communication channel is a very relevant goal~\cite{Flamini2019}. As routinely demonstrated in optical experiments, the utilization of the photonic information space can be maximized in very by playing with the time encoding~\cite{yokoyama2013}, spatiotemporal modes~\cite{armstrong2012, Richardson2013}, carrier frequencies and wavelengths~\cite{Roslund2014, Ciurana2014, Wengerowsky2018} and any other photonic degree of freedom.

In this work we explore two information multiplexing techniques that can be applied to microwave waveguide QED quantum links with single-photon wavepacket shaping techniques~\cite{Pechal2013, forn-diaz2017, kannan2023, Yang2023}. These techniques are explored and tested in a distributed architecture with two distant nodes connected by a waveguide QED quantum link, in line with recent setups~\cite{Pechal2013, Kurpiers2017, Magnard2020, Storz2023, Qiu2023, Grebel2024}. Our first result is to demonstrate a spatio-temporal mode multiplexing protocol, designing an explicit control to generate a large basis of orthogonal modes that is compatible with existing single-photon shaping setups~\cite{Pechal2013, Yang2023}. While this technique enlarges the accessible space for the construction of large photonic states~\cite{Schwartz2016, Ferreira2022} and generic quantum resources, we conclude that it is not possible to operate on two orthogonal photons that share the same carrier frequency due to cross-talk effects.

Motivated by this, we investigate frequency multiplexing protocols with multiple emitters and receivers operating at different wavelengths. A detailed study with 2+2 qubits sharing a common quantum link reveals that it is possible to simultaneously transfer two photonic qubits with large fidelity. Only a relatively small frequency separation, determined by the photon bandwidth, is required to reach the excellent fidelities of single-photon state transfer~\cite{Penas2023, Penas2022}. A pessimistic extrapolation of this to multiple emitters estimates a waveguide's information capacity, understood as the bandwidth requirements to transmit a given number of qubits below a given overall infidelity. These estimates hint that current setups support transferring tens of multiplexed qubits with minimal coherent errors at levels compatible with fault-tolerant computation requirements.

The structure of the article is as follows. In Sec.~\ref{sec:Model} we present the setup and introduce the one and two excitations Ans{\"a}tze that will be used for the simulations. In Sec.~\ref{sec:mode_multip} we describe  mode multiplexing, provide the theoretical details for the generation of the desired family of orthogonal wavepackets, and present numerical results of quantum state transfer protocols. In Sec.\ref{sec:freq_multip} frequency multiplexing is discussed, including detailed numerical simulations of quantum state transfers involving two photons to analyze its performance, and estimate its scaling with a growing number of multiplexed photons and emitters. Finally, in Sec.~\ref{sec:conclusions} we summarize the main contributions of the present work and provide an outlook for potential future applications.

\section{Model} \label{sec:Model}
This article aims to study and enhance the information capacity of a waveguide-QED quantum link between two quantum processors. Our theoretical model for the link consists of two quantum nodes connected to the ends of an open waveguide. Each node or station contains a quantum registers with one or more qubits, which are separately coupled to the waveguide by individual quantum filters. The following subsections describe the quantum optical models for this setup, as well as the exact wavefunctions that describe the single- and two-photon dynamics.

\subsection{Setup}\label{sec:setup}

\begin{figure}
    \centering
\includegraphics[width=\columnwidth]{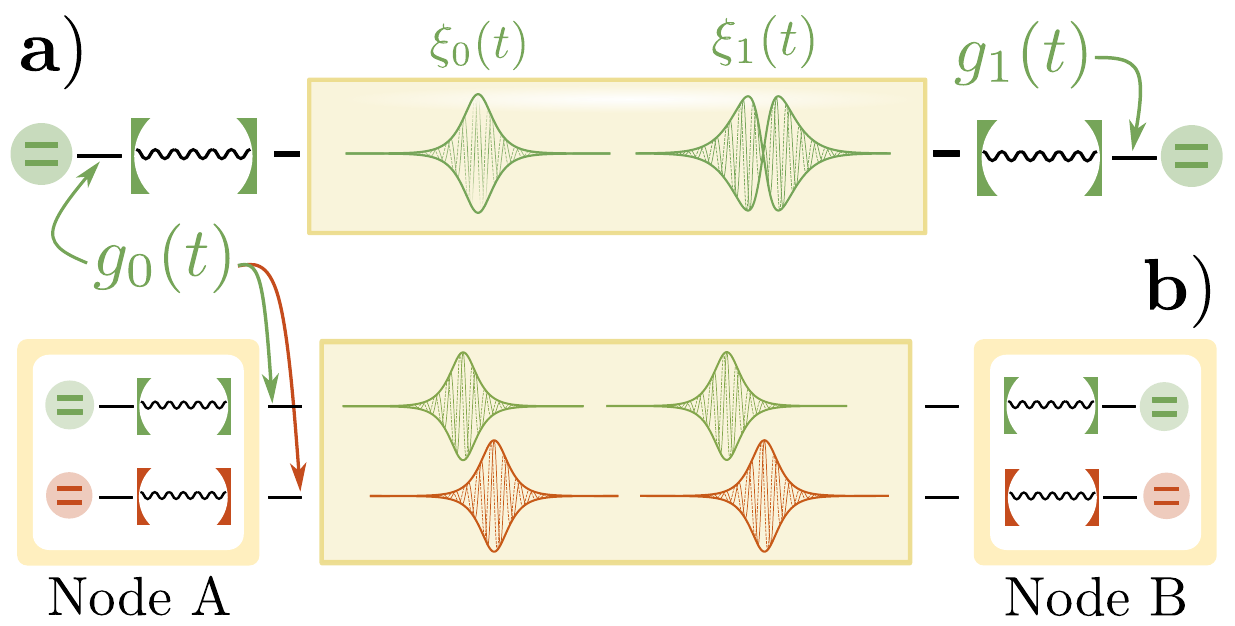}
\caption{Schematic representation of the multiplexation protocols considered in this article, namely, mode (a) and frequency multiplexation (b) across a waveguide connecting two nodes. Mode multiplexation is illustrated in (a), where one can apply either $g_0(t)$ or $g_1(t)$ to produce a propagating photon, $\xi_0(t)$, or one that is orthogonal to it, $\xi_1(t)$. At the second node one can selectively choose to interact with one or the other also by choosing the adequate control. Panel (b) refers to frequency multiplexing, illustrated for nodes containing two qubit-resonator systems each, and chosen to be pairwise resonant, $\delta_1 = \delta_3$ (green qubits), and $\delta_2=\delta_4$ (red qubits). The photons travel concurrently and do not disturb each other because they have different central frequencies, exploiting the available bandwidth. Note that the emitters are not restricted to interact only with their corresponding resonant photons. A careful derivation of the controls allows us to couple qubits with photons that are detuned from them by properly tuning the time-dependent coupling $g(t)$ (see Sec.~\ref{sec:mode_multip}  and~\ref{sec:freq_multip} for details).} 
\label{fig1}
\end{figure}

Our general waveguide-QED link comprises multiple emitters connected to a waveguide by transfer resonators that act as frequency filters. While variations of this setup have been used in state-of-the-art experiments with superconducting circuits~\cite{Eichler2012, Pechal2013, Zeytinoglu2015, Kurpiers2017, Magnard2020, Storz2023, Qiu2023}, we extend this framework to account for several emitters in each of the two nodes (cf. Fig~\ref{fig1}). The Hamiltonian for $N$ emitters, $N$ filters and the waveguide reads as follows
\begin{align}\label{eq:H}
    H &=H_{\rm WG}+\sum_{j=1}^N H_j+H_{j{\rm -WG}},\\
    H_{\rm WG}&=\sum_{k} \omega(k)b^\dagger_k b_k,\\
    H_{j}&=\delta_j \sigma^+_j\sigma_j^-+\omega_{{\rm R}j} a^\dagger_j a_j+g_j(t)\left(\sigma^+_ja_j +{\rm H.c.}\right), \\
    H_{j{\rm -WG}} & =\sum_k G_{k,j} \left(b_k^\dagger a_j+{\rm H.c.}\right).
\end{align}
The operators $\sigma^+_j$, $a^\dagger_j$, $b^\dagger_k$ respectively describe the creation of excitations in the qubit, filter and waveguide degrees of freedom. The energy of the qubits is denoted by $\delta_j$ and is approximately resonant to the frequency of the resonator it interacts with $\delta_j\simeq\omega_{Rj}$. Each qubit is coupled to a different filter, which act as bridges of information in and out of the waveguide. The qubit-cavity coupling amplitude $g_j(t)\in\mathbb{C}$ can be controlled both in amplitude and phase, a feature that can be achieved experimentally by using tuneable couplers~\cite{Yang2023} or mediating the interaction via the superconducting circuit states outside the computational basis~\cite{Pechal2013, Zeytinoglu2015}.

The waveguide dispersion relation $\omega(k)$ and the filter-waveguide coupling amplitudes $G_{k,j}$ are motivated by recent experiments. In particular, for the waveguide we assume a superconducting WR90 microwave waveguide of length $l_{\rm WG}$ (as employed in Refs.~\cite{Kurpiers2017,Magnard2020}). This waveguide features a non-linear dispersion relation of the form $\omega(k)=c\sqrt{(\pi/l_{1})^2+ k^2}$~\cite{Pozar} operating in the X-band, and where $c$ and $l_1$ refer to the speed of light and its cross-section dimension, $l_1=2.286$ cm. The wavenumber of these modes is given by $k_m=m\pi/l_{\rm WG}$ with $m=0,1,\ldots$ denoting the mode number. The amount of waveguide modes $N_{\rm WG}$ is chosen such that it ensures the convergence of numerical simulations. The resonator-waveguide coupling follows an Ohmic, $G_{k_m,j}=(-1)^{m(j-1)}\sqrt{\kappa_j v_g \omega(k_m)/(2\omega_{{\rm R}_j}l_{\rm WG})}$, where $v_g=d\omega(k)/dk$ is the group velocity and $\kappa_j$ is the decay rate of the $j$-th filter---i.e. the speed at which photons are released from the resonator and thus the largest bandwidth of our engineered photons.

\subsection{Single- and two- excitation state}

The Hamiltonian~\eqref{eq:H} commutes with the number of excitations, $\hat{n}=\sum_j \sigma^+_j\sigma_j+a_j^\dagger a_j + \sum_k b^\dagger_kb_k$. Thus, $\hat{n}$ is a conserved quantity in the dynamics and the Hamiltonian $H$ is block diagonal, with one sparse matrix describing the evolution in each sector of the Hilbert space with fixed $\hat{n}\in\{0,1,2,\ldots\}.$

If the initial state only contains one excitation, the whole evolution can be described by an exact wavefunction that is a superposition of one excited qubit or one photon living in either the filters or the waveguide
\begin{equation}\label{eq:ansatz_1exct}
  \ket{\Psi(t)} = \left[\sum_{j}\left(q_j(t)\sigma_j^++c_j(t)a_j^\dagger\right)+\sum_k \psi_k(t)b_k^\dagger \right]\ket{\mathbf{0}}.
\end{equation}
The coefficients $q_j(t), c_j(t), \psi_k(t)$, are complex numbers denoting the probability amplitudes of exciting a qubit, resonator or waveguide mode, respectively, starting from the vacuum state is denoted by $\ket{\mathbf{0}}$. The evolution of state~\eqref{eq:ansatz_1exct} according to Hamiltonian~\eqref{eq:H} is dictated by a set of coupled ordinary differential equations for the probability amplitudes
\begin{align}\label{eq:eom1}
    i\dot{q}_j(t)&=\delta_j q_j(t)+g_j(t) c_j(t),\\ \label{eq:eom2}
    i\dot{c}_j(t)&=\omega_{{\rm R}j} c_j(t)+g^*_j(t) q_j(t)+\sum_k G_{k,j}\psi_k(t),\\ \label{eq:eom3}
    i\dot{\psi}_k(t)&=\omega(k) \psi_k(t)+\sum_{j}G_{k,j}c_j(t).
\end{align}

The analysis of multiplexing strategies requires us to go beyond the single excitation limit. We stop at the two excitation limit, which is the next one that is still numerically tractable. The Schwinger model for this wavefunction is
\begin{align}\label{eq:ansatz_2exct} \nonumber
  \ket{\Psi(t)} = \bigg[\sum_{i \neq j}q_{i,j}(t)\sigma_i^+\sigma_j^+ + \sum_{i,j}c_{i,j}(t)a_i^\dagger a_j^\dagger+\sum_{mn} \psi_{m,n}(t)b_m^\dagger b_n^\dagger \\  + \sum_{i,j} qc_{i,j}(t)\sigma_i^+ a_j^\dagger + \sum_{i,m} q\psi_{i,m}(t)\sigma_i^+ b_m^\dagger + \sum_{i,m} c\psi_{i,m}(t) a_i^\dagger b_m^\dagger \bigg]\ket{\mathbf{0}}.
\end{align}
Following the single excitation notation, $q_{i,j}(t), c_{i,j}(t)$ and $\psi_{m,n}(t)$  respectively describe two excitations in the qubits, the resonators or the waveguide. The wavefunction also includes the possibility of excitations coexisting in the qubit and a filter $qc_{i,j}(t)$, the qubit and a waveguide mode $q\psi_{i,m}(t)$, and a combination of filter and waveguide modes $c\psi_{i,m}(t)$. The necessary restriction that no qubit alone houses more than one excitation is fixed by the condition $i \neq j$ in the first sum. The evolution of state~\eqref{eq:ansatz_2exct} under Hamiltonian \eqref{eq:H} produces a set of coupled differential equations analogous to Eqs.~\eqref{eq:eom1}-\eqref{eq:eom3}.

The following study relies on numerically exact solutions to the problems~\eqref{eq:eom1} and \eqref{eq:ansatz_2exct}. Both problems are tractable and can be simulated with moderate resources up to thousands of modes. In all cases we analyze protocols that operate under different controls $g_j(t)$ to achieve multiplexed state transfer from initial states with one or two excited qubits, to final states in which the information has been transferred to a different node. The derivation of the time-dependent couplings $g_j(t)$ will vary according to the multiplexing strategy and will be discussed in Sec.~\ref{sec:orthogonal_photons}, as well as in Appendix~\ref{app:orthogonal_gs}.

\section{Mode multiplexing}\label{sec:mode_multip}

Our first multiplexation strategy adds a new spatio-temporal degree of freedom, which is the waveform of the propagating photons in the waveguide. Thus, on top of a qubit degree of freedom---having zero or one photon in a waveguide---, a new quantum degree of freedom is the label of the photon shape within an orthonormal family of modes [cf. Fig.~\ref{fig1}a]. This strategy is only possible if we calibrate the controls $g_j(t)$ to selectively emit and absorb an orthonormal basis of functions, $\xi_0(t)$, $\xi_1(t)$, etc. The new controls must be selective, so that if the $j$-th qubit is designed to absorb a given wavepacket shape, it will perfectly reject photons in orthogonal mode. As we will see below, having more than one possible mode requires departing from conventional wavepacket shaping techniques~\cite{Vogell2017,Magnard2020,Penas2022,Pichler2017}, which assume real-valued positive wavepackets, to use more sophisticated designs.

\subsection{Engineering orthogonal wavepackets}
\label{sec:orthogonal_photons}

We are interested in producing a discrete family of photon wavepackets, which we denote $\{\xi_n(t)\}$ with a discrete label denoting the mode index $n=0,1,\ldots$  These photonic modes from an orthonormal basis
\begin{align} \label{eq:ortho_condition}
    \langle \xi_n(t),\xi_m(t)\rangle = \int_{-\infty}^{\infty}dt \xi_n^*(t)\xi_m(t)=\delta_{n,m}.
\end{align}
For this to happen it is mandatory that one or more modes have non-trivial phase profiles---this may be associated to $\xi_n(t)$ having zeros and changing sign at specific times, or to a more sophisticated phase front. This requirement poses a challenge to the conventional wavepacket shaping techniques, which assume real-valued functions for wavepackets and controls along the calculations. 
Fortunately, this is not an intrinsic limitation, and we can explore more generic complex controls $g(t)\equiv |g(t)|e^{-i\varphi(t)}$, as enabled, for instance, by state-of-the-art qubit-cavity coupling schemes~\cite{Pechal2013}.


As in standard wavepacket shaping, the design of $g(t)$ is the solution of an inverse problem, in which the control is obtained by inverting the dynamical equations (cf. for a single excitation) for a single emitter creating a photon with a predefined shape. To make the inversion tractable, we must replace the full dynamical equations Eqs.~\eqref{eq:eom1}-\eqref{eq:eom3} with an alternative model that works in the Markov limit, and which is analogous to standard input-output theory~\cite{Gardiner1985,morin2019deterministic, gorshkov2007photon}. The effective model is obtained by identifying the coupling of the resonator to the waveguide $\sum G_{k,j}\psi_k(t)$ as a Markovian decay channel with rate $\kappa$. For details regarding the validity of the approximation in these systems with time dependent couplings we refer to the Appendix B in Ref.~\cite{Penas2023}. The resulting model in the single excitation limit is
\begin{align} \label{eq:effective_Markov_model}
    \dot{q}(t)&=-i\omega_R q -i |g(t)|e^{+i\varphi(t)}c(t),\\
    \dot{c}(t)&=-i\omega_R c -i |g(t)|e^{-i\varphi(t)}q(t)-\kappa c(t)/2,
\end{align}
together with the output relation $\xi(t) = i\sqrt{\kappa}c(t)$.

These expressions can be transformed into a relation between the qubit excited probability to the probability distribution of photon wavepacket
\begin{align} \label{eq:modulus_q}
|q(t)|^2=|q(t_0)|^2-\frac{|\xi(t)|^2}{\kappa}-\int_{t_0}^{t}d\tau |\xi(\tau)|^2.
\end{align}
Note that Eq.~\eqref{eq:modulus_q} limits the type of wavefronts that can be generated, by bounding the logarithmic derivative of the wavepacket $\xi(t)$. In other words, if the photon profile raises (or falls) faster than an exponential, then the r.h.s of \eqref{eq:modulus_q} becomes negative and the model breaks down.

For convenience we move to the rotating frame of the carrier frequency $\omega_R$, so that when the photon envelope is a real-valued positive function  $\xi(t) = |\xi(t)|$, the control $g(t)$ is also a non-negative function $g(t)=|g(t)|$ obtained explicitly from~\eqref{eq:modulus_q} and \eqref{eq:effective_Markov_model}, i.e. $\varphi(t)=0$. As shown in Ref.~\cite{Penas2023}, more general photon wavepackets, with sign changes and complex phase profiles that correct deficiencies in the waveguide are possible. However, it is even possible to find explicit, closed expressions for the control $g(t)$ in terms of a generic photon wavepacket $\xi(t)$. The derivation of these formulas requires us to separate the photon waveform into amplitude and phase 
\begin{align}
    \xi(t)=f(t)e^{-i\theta(t)},
\end{align}
with $f(t),\theta(t) \in\mathbb{R}$, with the caveat that $f(t)$ may take negative values~\footnote{Removing the sign dependence of $\xi(t)$ from $\theta(t)$  eases the calculations, as otherwise $\theta(t)$ may undergo discontinuous jumps whenever $\xi(t)$ changes sign.}. Next,  defining $F(t)=\int_{t_0}^t d\tau |\xi(\tau)|^2=\int_{t_0}^{t} d\tau f^2(\tau)
$ allows rewriting Eq.~\eqref{eq:modulus_q} as $|q(t)|^2 = 1- f^2(t) / \kappa -F(t)$. Taking the derivative of this last expression and substituting into \eqref{eq:effective_Markov_model} produces a closed expression for the modulus of the control only in terms of wavepacket variables,
\begin{align}\label{eq:gtmod}
    |g(t)|=\sqrt{\frac{(\dot{f}(t)+\kappa/2 \ f(t))^2+(\dot{\theta}(t) \ f(t))^2}{\kappa(1-F(t))-f^2(t)}}.
\end{align}

Finally, one can relate the phase of the qubit amplitude ${\rm arg}\{q(t)\}$ with that of the wavepacket $\theta(t)$ (see App. A in~\cite{Penas2023} for details) and obtain an expression in an integral form,
\begin{align}\label{eq:phasedot}
    \varphi(t)=\pi&+\theta(t)+\phi_f(t)+{\rm atan}\left(\frac{\dot{\theta}(t)f^2(t)}{-(\dot{f}(t)f(t)+\kappa/2 f^2(t))}\right)\nonumber\\&-\int_{t_0}^td\tau \frac{\dot{\theta}(\tau)f^2(\tau)}{\kappa(1-F(\tau))-f^2(\tau)},
    \end{align}
where $\phi_f(t)$ accounts for the sign change of $f(t)$, namely, $\phi_f(t)=0$ or $\pi$ when $f(t)\geq 0$ or $f(t)<0$, respectively, i.e. $\xi(t)=|f(t)|e^{-i\theta(t)-i\phi_f(t)}$.

Equations~\eqref{eq:gtmod} and \eqref{eq:phasedot} provide us the control $g(t)$ that engineers a photon in mode $\xi(t)$. As in usual state-transfer protocols~\cite{Cirac1996}, the time-reversed control $g(-t)$ is the one that enables perfectly absorbing the generated wavepacket $\xi(t)$ in the receiver station.

This design includes a specific application, which is shifting the wavepacket's carrier frequency from the resonator $\omega_R$ to a slightly different value $\omega_R+\delta\omega$. In the rotating frame, this means a wavepacket
\begin{equation}
  \xi_\delta(t) = \xi(t) e^{-i\delta\omega t}, 
\end{equation}
obtained when $\theta(t)=\delta \omega t$ and thus $\dot{\theta}(t) = \delta\omega$. In this case, $\varphi(t)$ becomes a small chirp that compensates for the frequency change. This chirping mechanism is a tool that has been used experimentally to compensate for experimental deviations between the emitter and receiver qubits~\cite{Magnard2020, Storz2023}. The experimentally obtained chirp can now be rigorously derived from ab initio principles. It can also be generalized to other applications, such as addressing photons from one emitter to various receiving nodes that differ in frequency.

\subsection{Orthogonal mode state transfer}
 \begin{figure}
    \centering
    \includegraphics[width=\columnwidth]{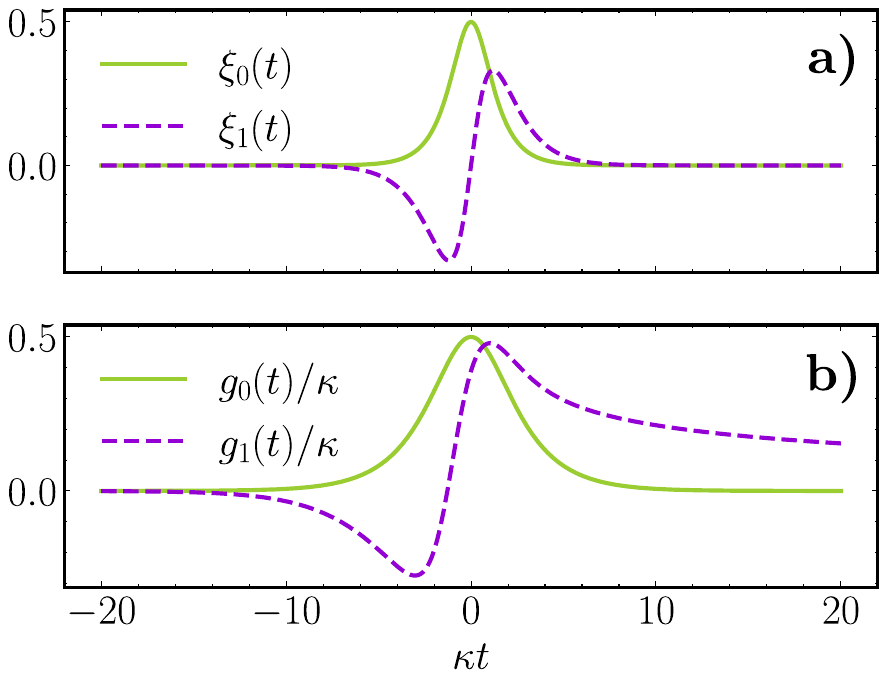}
    \caption{Panel a) shows the shape of the first two mutually orthogonal photons from the family $ \xi_{n}(t) \propto \sech(t) \times t^n$, i.e. $\xi_0(t)$ and $\xi_1(t)$. Panel b) shows the corresponding controls to produce them, $g_0(t)$ and $g_1(t)$, respectively. See Appendix \ref{app:orthogonal_gs}, Eqs.~\eqref{eq:xi0_ap}-\eqref{eq:g1_ap}.}
    \label{fig:orthogonal_controls}
\end{figure}

Let us now benchmark numerically the design of a single-photon state-transfer protocol that makes use of a family of two orthogonal modes, $\xi_0(t)$ and $\xi_1(t)$, respectively created by the controls $g_0(t)$ and $g_1(t)$. The study will confirm that using the same control on both nodes allows transferring one qubit of information between the two nodes that are connected to the ends of the waveguide. It should also confirm that using different controls allows the receiver to fully reject the incoming photon with 100$\%$ probability.

We construct the orthonormal family of photon wavepackets starting from a sech-pulse
\begin{equation}
  \label{eq:sech-pulse}
  \xi_0(t) = \sqrt{\frac{\kappa}{4}}\sech(\kappa t/2) 
\end{equation}
that is a commonly used photon in the literature~\cite{Kurpiers2017,Magnard2020,Axline2018,Campagne2018}. As first orthogonal mode we use the antisymmetric waveform with one zero
\begin{align}\label{eq:first_ortho_photon}
 \xi_1(t)=\sqrt{\frac{3\kappa^3}{4\pi^2}}\sech(\kappa t/2) t 
\end{align}
A larger family of sech-orthogonal photons can be constructed by Gram-Schmidt orthogonalization (see Appendix~\ref{app:orthogonal_gs}), but for this study two modes suffice.

\begin{figure}
    \centering
\includegraphics[width=\columnwidth]{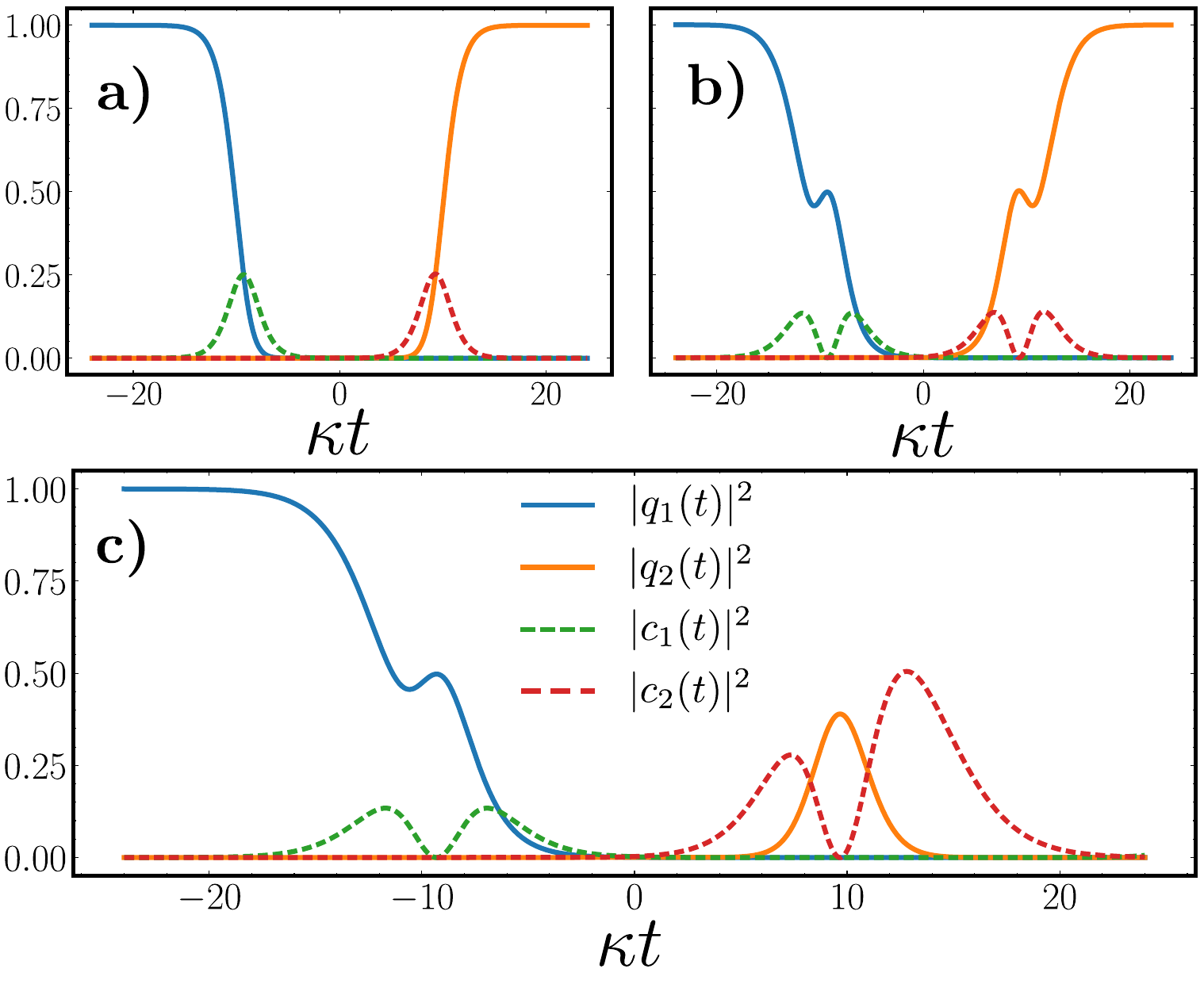}
\caption{Quantum state transfer employing orthogonal wavepackets, showing the evolution of the populations of the qubits and transfer resonators. Panels a) and b) show a standard emission absorption protocol using the same controls at both ends, namely $g_0(t)$ in (a) and $g_1(t)$ in (b), resulting in a transfer efficiency in both cases well over $99\%$. Panel c) shows the dynamics when a photon $f_1(t)$ is emitted using $g_1(t)$,  and apply an orthogonal control $g_0(t)$ at the receiver end. As the receiver is designed to absorb $f_0(t)$, which is orthogonal to the incoming photon $f_1(t)$, the transfer efficiency vanishes, i.e.  $\approx 10^{-5}$. The simulation parameters are: $\delta_{1,2}=\omega_{1,2} = 2\pi \times 8.9$ GHz, $N_\text{WG} = 300$, $l_\mathrm{WG}  =30$ m, and $\kappa_{1,2}= 2\pi \times 20 $MHz.} 
\label{fig2}
\end{figure}


The controls that create these two wavepackets can be obtained analytically by plugging $f_0(t)=\xi_0(t)$ and $f_1(t)=\xi_1(t)$ into Eqs.~\eqref{eq:gtmod} and~\eqref{eq:phasedot}, with $\theta_0=\theta_1=0$. This results in two functions explicitly given in App.~\ref{app:orthogonal_gs}. We can compare the shape of the wavepackets $f_0(t)$, $f_1(t)$ [cf. Fig.~\ref{fig:orthogonal_controls}a] to the controls that create them $g_0(t)$, $g_1(t)$ [cf. Fig.~\ref{fig:orthogonal_controls}b]. Note how $f_1(t)$ changes sign: such wavepacket cannot be produced by a non-negative coupling $g_1(t) = |g_1(t)|$, but our generalization still creates a smooth and real control, where the resulting phase $\varphi(t)=0,\pi$ just reflects a change of sign in $g(t)$. Importantly, such change of sign stems from $f(t)\geq 0$ and $f(t)<0$, i.e. $\phi_f(t)$, as well as from the quantity $\dot{f}(t)f(t)+\kappa/2 f^2(t)$ being positive or negative (cf. Eq.~\eqref{eq:phasedot}). 

Fig.~\ref{fig2} shows a numerical simulation of the single-qubit state-transfer protocol with two orthogonal modes, using an experimentally motivated parameterization of the model in Sec.~\ref{sec:setup}. The two top panels show how applying the same control $g(t)$ at both ends leads to an almost perfect absorption or transfer efficiency~\footnote{Note that the absorption is not strictly perfect due to distortion effects of the travelling wavepacket, as studied in Ref.~\cite{Penas2022}. The correction techniques discussed in Ref.~\cite{Penas2023} may improve the outcome, but we decided to avoid them, thus allowing a straight comparison of with the single-mode strategy.}, measured by the probability of transferring all the excited population of the emitter qubit at one end of the link, $|q_1(0)|^2=1$, to the receiver qubit at the end of the protocol, $|q_2(T)|^2$.  The numerical simulations from Fig.~\ref{fig2}c confirms the selective nature of the controls, by which a control designed to absorb a given photon mode $\xi_0(t)$ fully rejects an incoming photon in mode $\xi_1(t)$. Note that, although not shown, this situation is equivalent to reverse scenario, i.e. emitting with $g_0(t)$ and absorbing with  $g_1(t)$ and can be extended to arbitrary mode functions, $\xi_n(t)$, see Appendix~\ref{app:orthogonal_gs}.

It is important to remark that the rejected photon experiences a strong distortion by the interaction with the second qubit. While we don't explicitly show the wavefront in the plot, both the initial wavepacket $|\xi_1(t)|^2\propto \kappa_1|c_1(t)|^2$ and the distorted one $|\xi_1(t)'|^2\propto \kappa_2|c_2(t)|^2$ can be deduced, by input-output theory, from the excitation profiles of the filters. The asymmetric distortion of the right-hand node is remarkedly different from the original antisymmetric profile [cf. Fig.~\ref{fig2}c, red thick dashed line]. As discussed in detail in Appendix~\ref{app:scattering}, this distortion can be fully explained by the scattering of the photon wavepacket by the resonant qubit-filter system. The photon, while rejected, still interacts with the qubit during the reflection process, experiencing a phase distortion that is similar to the ones studied in Ref.~\cite{Penas2022}. This distortion can be compensated in later interactions with the rejected photon, using a dynamical protocol such as the one in Ref.~\cite{Penas2023}. This makes us confident to state that mode multiplexing can be used not only in state transfer, but also in generic protocols for high-dimensional photonic state generation~\cite{Ferreira2022}.

However, while single-photon mode multiplexing works, we have found it impossible to simultaneously generate two photons in two orthogonal modes. A setup that involves two resonant emitters, coupled with different controls $g_0(t)$ and $g_1(t)$, still gives rise to a strong cross-talk and a deterioration of the photons produced. The natural solution to this problem is to place the emitters in different regions of the spectrum, to suppress their interaction. However, this gives rise to a different type of multiplexing strategy, frequency-type multiplexing, which is discussed in the following section.

\section{Frequency multiplexing} \label{sec:freq_multip}
Frequency or wavelength multiplexing refers to the transmission of multiple qubits, each one encoded in a separate photon with a unique carrier frequency. An illustration of the idea can be found in Figure~\ref{fig1}b, where the different colors depict photons at different frequencies travelling simultaneously through the waveguide. Frequency and mode multiplexing are not incompatible and may be combined in a single setup.

In this section we explore the performance of frequency multiplexing in microwave experiments with realistic microwave guide. Thestate-of-the-art experiments we reference in our study (e.g., Refs.~\cite{Kurpiers2017,Magnard2020}) operate mainly in the X-band , with a 4 GHz bandwidth (i.e. $\sim 8-12$ GHz) that may potentially host multiple photons, each with a width in the range of a few megahertz.

As a minimal model to study the simultaneous generation and transfer of two flying qubits, we consider the setup in Figure~\ref{fig1}b, with two nodes, each hosting two qubits, that are paired in frequency. Following the Hamiltonian from Sect.~\ref{sec:setup}, where qubits 1 and 2 belong to the first node and 3 and 4 belong to the second one, the frequencies are matched for the qubits and resonators $\delta_1=\omega_{R1}=\delta_3=\omega_{R3}$ and $\delta_2=\omega_{R2}=\delta_3=\omega_{R3}$. The study focuses on the single and two excitations subspaces, which fully captures the state transfer of up to two qubits. The evolution in these subspaces is exactly captured by wavefunctions~\eqref{eq:ansatz_1exct} and \eqref{eq:ansatz_2exct}.

The goal of the study is to analyze the errors made by transmitting two qubits at different frequencies. Qualitatively, we expect two sources of error in this process: there will be an error due to the imperfect distinguishability of the photons associated to the individual qubits, and there will be another type of errors associated to the cross-talk between qubits and resonators operating at different frequencies. We expect that both errors will die off as the frequency separation exceeds the bandwidth of each individual photon  $\Delta_{12} \equiv |\delta_1-\delta_2| \gg \kappa$, a reason that motivates us to use the same controls $g(t)$ as prescribed by individual state transfer processes. However, this statement and the behavior of different errors must be studied more carefully in general conditions.

\subsection{Emitters cross-talk}
As soon as we connect multiple qubits and filters to the same environment, the waveguide, there is a probability that these quantum objects may influence each other, even if their resonance frequencies are far apart. This mutual influence, generally called cross-talk, can manifest in an actual exchange of energy between quantum objects---e.g., a qubit living in site $1$ may jump to sites $2$ and viceversa---or in the change of intrinsic properties of the emitters--e.g., renormalization of frequencies, decay rates, etc.

To study the impact of cross-talk in the limit of large frequency separation we developed an effective model that describe the mutual interaction among the two quantum filters in a single node. As showin in App.~\ref{app:mutual_influence}, the effective Hamiltonian reads
\begin{equation}\label{eq:H_exchange_terms}
    H_\text{eff} = \sum_i (\omega_{Ri} + \delta \omega_{Ri})a^{\dagger}_i a_i  +  \tilde{G}\left( a_1 a_2^\dagger  + a_1^\dagger a_2\right).
\end{equation}
This Hamiltonian exemplifies the two effects mentioned before: the natural frequency of the resonators is changed, $\delta \omega_{Ri}$, and the two filters acquire a coherent coupling $\tilde{G}$. The shift in energy of the resonators will affect the carrier frequency of the flying qubits, thereby influencing the probability that flying qubits can be captured using our controls. Second, the exchange term $G$ enables the possibility that qubits 1 and 2 swap information, scrambling the transmitted information.

An analysis of the effective Hamiltonian \eqref{eq:H_exchange_terms} reveals that exchange terms $a_i a_j^\dagger$ dominate the dynamics in the limit in which mode separation is smaller than the filters' bandwidth  $|\omega_{R2} -\omega_{R1}| \ll \kappa $. Interestingly, the infidelity in this regime can still be estimated by the frequency overlap between the injected photons, as this quantity describes the indistinguishability in frequencies and is related to the probability that photons from one wavepacket can tunnel into the opposite resonator. A quantitative analysis, supported by numerical simulations is shown in Sect.~\ref{sec:2phot}.

A more interesting limit is one in which the frequency separation is larger than the spectral width of the resonators, i.e. $|\omega_{R1}-\omega_{R2}| \gg \kappa$. We are interested in learning whether we can recover the ideal state-transfer fidelities of single-photon processes. This limit is captured by a Magnus expansion on \eqref{eq:H_exchange_terms}
\begin{align}\label{eq:eff_eff_Hamiltonian_main}
    \tilde{H}_\text{eff} = \sum_i C_i a_i^{\dagger} a_i; \; \; \text{with} \; \; C_i = (-1)^{i+1} \frac{\tilde{G}^2}{\omega^D_{R2}-\omega^D_{R1}},
\end{align}
where we again refer to Appendix \ref{app:mutual_influence} for the details. The effective Hamiltonian~\eqref{eq:eff_eff_Hamiltonian_main} indicates that even when the detuning between the resonators is large, there is a mutual influence that modifies the natural frequency of the filters.

To verify this prediction and better understand the cross-talk during the state-transfer dynamics, we have run spectroscopy-like experiments exploring the renormalization of the filters' properties. These experiments copy the setup from Fig.~\ref{fig2}b, but only excite qubit 1. The qubit-resonator coupling $g_1(t) = \kappa/2 \; \sech(\kappa t/ 2)$ is designed to fully transduce the qubit into a flying photon under ideal conditions. By tuning the frequency of the qubit $\delta_1$ and varying the parameter $\kappa$, we can analyze the filter's intrinsic resonance and its renormalized decay rate. As discussed in App.~\ref{app:mutual_influence}, we find that both quantities are affected by the cross-talk with the neighboring resonators, and the change in frequency and bandwidth accurately follow the theoretical predictions.

The results from this study are very promising. First, the spectroscopy method allows us to determine the dressed properties of the filter in a way that is compatible with what can be done experimentally---where we only have access to the renormalized properties. Second, while the properties of the filter are affected by cross-talk, the simulations reveal that these changes only influence the photon wavepacket---i.e., carrier frequency and bandwidth---and that perfect state transfer is still possible, provided we redesign the controls $g_i(t)$ to account for the changes in the flying qubits. The important remaining question is whether these effective renormalizations also describe the dynamics with multiple flying qubits.

\subsection{Multiplexed state transfer tomography}
The simplest setup we can use to explore frequency multiplexing, sketched in Fig.~\ref{fig1}b, consists of two pairs of qubits, each pairwise resonant $\delta_1=\delta_3$ and $\delta_2=\delta_4$, with qubits 1 and 2 in node A, and qubits 3 and 4 in node B. In the ideal situation, the state transfer process will map any state of each qubit from node A to the corresponding qubit in node B, satisfying
\begin{align}\label{eq:ST2exct}
U_\text{ideal}\left[ (\alpha \ket{0}_1+\beta\ket{1}_1) (\delta \ket{0}_2+\gamma\ket{1}_2) \otimes \ket{0}_3 \ket{0}_4\otimes\ket{\textbf{0}} \right]  \\
= \ket{0}_1 \ket{0}_2 \otimes(\alpha \ket{0}_3+\beta\ket{1}_3) (\delta \ket{0}_4+\gamma\ket{1}_4)\otimes\ket{\textbf{0}}, \nonumber
\end{align}
for any set of normalized complex weights $|\alpha|^2+|\beta|^2=|\gamma|^2+|\delta|^2=1$. Here, the first four indices correspond to the qubits, and the fifth index (bold font) refers to the bosonic degrees of freedom, which remain in the vacuum state at the beginning and end of the transformation.

A protocol that fulfills this equation completely maps the information stored in the $d=4$ states of the first node's computational basis $\ket{i,A}\in \{\ket{00 \; 00 \; \mathbf{0}}$, $\ket{01 \; 00 \; \mathbf{0}}$, $\ket{10 \; 00 \; \mathbf{0}}$, $\ket{11 \; 00 \; \mathbf{0}}\}$, to the four corresponding computational states of the second node $\ket{i,B}\in \{\ket{00 \; 00 \; \mathbf{0}}$, $\ket{00 \; 01 \; \mathbf{0}}$, $\ket{01 \; 00 \; \mathbf{0}}$, $\ket{11 \; 00 \; \mathbf{0}}\}$. The following isometry captures this ideal transformation
\begin{equation}
  U_\text{ideal} = \sum_{i}^{4} \ket{i,B}\!\bra{i,A}.
\end{equation}
To characterize the experimental setup, we only need to reconstruct the isometry that maps the four input states $\ket{i, A}$ to other states of the quantum links and nodes
\begin{equation}
  U_\text{real} = \sum_{i,i'=1}^4 A_{i'i}\ket{i'}\!\bra{i} + B.
\end{equation}
The coefficients $A$ will not be the identity, and there will be leakage $B$ to other states in the Hilbert space--e.g., photons that are not absorbed in the receiver qubits and wander through the waveguide or return to the original station.

The errors in the state transfer operation may be quantified using either the entanglement fidelity or the average fidelity, which respectively are~\cite{Nielsen2002, Pedersen2007}
\begin{align}
  \mathcal{F}_\text{ST}^{(e)} &= \left|\frac{1}{d} \text{Tr}\left( U_\text{ideal}^\dagger U_\text{real} \right)\right|^2 = \frac{1}{d^2} \left|\sum_{i=1}^{d} \braket{i,A| U_\text{ideal}^\dagger U_\text{real} |i,A}\right|^2,
                                \label{eq:stfidelity}\\
  \bar{\mathcal{F}}_\text{ST} &= \frac{d\mathcal{F}_\text{ST}^{(e)}+1}{d+1},
                                \notag
\end{align}
where $d=4$ is the dimension of the computational input and output spaces. We will use the entanglement fidelity $\mathcal{F}_\text{ST}^{(e)}$.

In actual simulations, these fidelities exhibit an oscillating behavior as a function of the frequency separation between qubits 1 and 2. This behavior can be explained by dynamical phases acquired by the different states in the computational basis due to the frequency difference among those states. However, these are phases that can be compensated for by applying local rotations on the individual qubits, either before or after the state transfer. Hence, we introduce the corrected two photon fidelities as those resulting from the maximization over those corrections
\begin{equation}\label{eq:ST2exctOpt}
\mathcal{F}_2 = \max_{\phi_0,\phi_1,\phi_2} \mathcal{F}_\text{ST}^{(e)}\left[e^{i\phi_0+i\phi_1\sigma^z_1+i\phi_2\sigma^z_2}U_\text{real}\right].
\end{equation}
Fortunately, this optimization is trivially performed by inferring the effective single- and two-qubit phases from $A_{ij}$.

Note that he following study will compare the 
accuracy of multiplexed two-qubit state transfer and the fidelity associated with single-qubit processes. It will also estimate the performance of setups with more than two flying qubits. $\mathcal{F}_1(\omega)$ indicates the fidelity of state transfer for one emitter on each side.


\subsection{Two-photon state transfer performance}\label{sec:2phot}

\begin{figure}
    \centering
\includegraphics[width=\columnwidth]{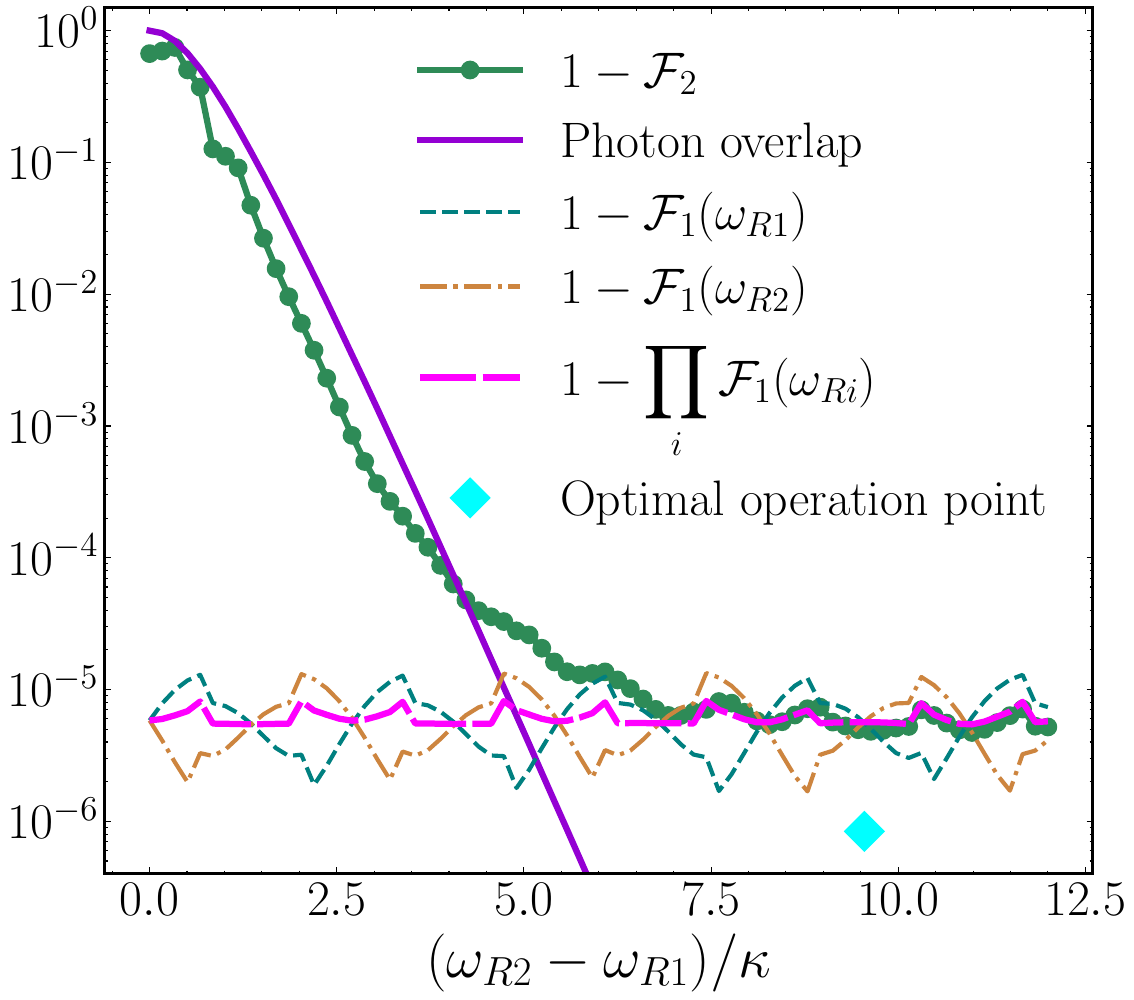}
\caption{Quantum state transfer protocol errors for two emitters as a function of their frequency separation. We plot the actual infidelity from Eq.~\eqref{eq:ST2exctOpt} (solid-dot) for multiplexed transfer; the single qubit infidelities for each emitter separately $1-\mathcal{F}(\omega_{R1,R2})$ (dashed and dot-dashed), and the lower bound for the multiplexed infidelity $1-\mathcal{F}(\omega_{R1})\mathcal{F}(\omega_{R2})$ (long dash). As an illustration, we also plot the optimized infidelity of multiplexed state transfer (big light-blue diamond) for a particular frequency separation. The solid line is a visual guide given by the photon wavepacket overlap~\eqref{eq:overlap_infidelity}. Qubit frequencies are symmetrically placed around $(\omega_{R1}+\omega_{R2})/2 = 2\pi \times 8.9$ GHz. The simulations were performed using $N_\text{WG} = 500$ modes, a $l_\text{WG} =15$ m long waveguide, a resonator bare decay rates of $\kappa =2\pi \times 10$ MHz, a hyperbolic secant control $g(t) = \kappa_\text{eff}/2 \text{sech} (\kappa_\text{eff} t / 2)$ and a protocol duration  $t \in (-35/\kappa, + 35/\kappa)$.}
\label{fig3}
\end{figure}

Let us analyze the performance of two-photon state transfer in the frequency multiplexed scenario. Our study will consider a waveguide that is $l_\mathrm{WG}=15$ meters long and which can be accurately modeled using $N_{\rm WG}=500$ frequency modes (see Sec.~\ref{sec:setup} for more details). Our analysis explores how the fidelity improves with the frequency separation of the flying qubits, starting from a resonant case $\omega_1=\omega_2$ until their separation significantly exceeds the bandwidth $|\omega_1-\omega_2| = 12\kappa$.

The bare frequencies of the qubits and filters that create these photons explore a moderate region of the waveguide bandwidth, using $\delta_i, \omega_{Ri} \in \left(2\pi \times 8.84, \;2\pi \times 8.96\right) \text{GHz}$  for the bare frequencies, and a bare decay rate for the resonators of $\kappa_i = 2 \pi \times 10$ MHz. More precisely, in consonance with the previous sections, the simulations fix the frequency of the bare resonators $\omega_{Ri}$ and use spectroscopy methods to calibrate the renormalized frequency and the effective decay rate$\kappa_\text{eff, i}$. These values are used then to optimally place the emitter qubit $\delta_i =\omega^D_i$ and to design the photon generation control $g_i(t)$. Note that, while the corrections to the original values are small---around $10^{2}$ kHz as discussed in App.~\ref{app:mutual_influence}---they significantly influence the state transfer fidelity.

Figure~\ref{fig3} shows the quantum operation's fidelity as the filter frequencies' frequency separation increases from 0 to $12 \kappa$. We observe a region dominated by the exchange of population between emitters, which happens when the separation is close to or below the individual photons' bandwidths. This region is followed by a plateau in which the physics reduces to two independent single-photon state transfer processes.

In the region where the detuning between the emitters is small, the infidelity of the state transfer process can be almost fully accounted for by the indistinguishability of the generated photons. Defining the overlap between two multiplexed modes centered on distinct carrier frequencies $\omega_1$ and $\omega_2$, i.e. $\xi(\omega_1)$ and $\xi(\omega_2)$, 
\begin{equation} \label{eq:overlap_infidelity}
\mathcal{I}_\text{overlap}(\omega_1, \omega_2) = \left|\bra{\xi(\omega_1)} \xi(\omega_2) \rangle\right|^2
\end{equation}
we find $1-\mathcal{F}_\text{ST}(\omega_{R1},\omega_{R2})\simeq I_\text{overlap}(\omega_{R1},\omega_{R2})$, as evidenced by the purple like in Fig.~\ref{fig3}. In the case of sech-like photons, this overlap admits an analytical expression that decays exponentially with the ratio $(\omega_{R1}-\omega_{R2})/\kappa$, further confirming the photon bandwidth as the natural scale to separate the emitters.

When the detuning between modes becomes much larger than $\kappa$, the emission and absorption of the two flying qubits become independent processes that behave as in the single excitation limit. The state transfer of each mode occurs at frequencies that deviate from the bare parameters $\omega_{R1,R2}$. Still, the qubit-assisted spectroscopy and tuning of $\delta_{i}$ enables fidelities compatible with the single-qubit studies. Indeed, as observed in Fig.~\ref{fig3}, for a frequency separation above $6\kappa$, the two-qubit state transfer fidelity approaches the product of fidelities of individual state transfer processes happening at frequencies $\omega_{R1}$ and $\omega_{R2}$. The oscillations found in this curve are also explained by the behavior of single-photon state transfer (cf. $\mathcal{F}(\omega_{R1})$ and $\mathcal{F}(\omega_{R2})$ in Fig.~\ref{fig3}). More precisely, in the limit of a 15 meter waveguide, a photon with a bandwidth of $\kappa\simeq 2\pi\times 10$ MHz can resolve the mode structure of the quantum link, and only when the filter $\omega_{R}$ coincides with one of the modes the transfer is optimal. To illustrate this, we show a point in Fig.~\ref{fig3} where we have optimized the placement of both resonators so that the two photons are perfectly in resonance with a waveguide mode.

\subsection{Scaling with N emitters}

\begin{figure}
    \centering
\includegraphics[width=\columnwidth]{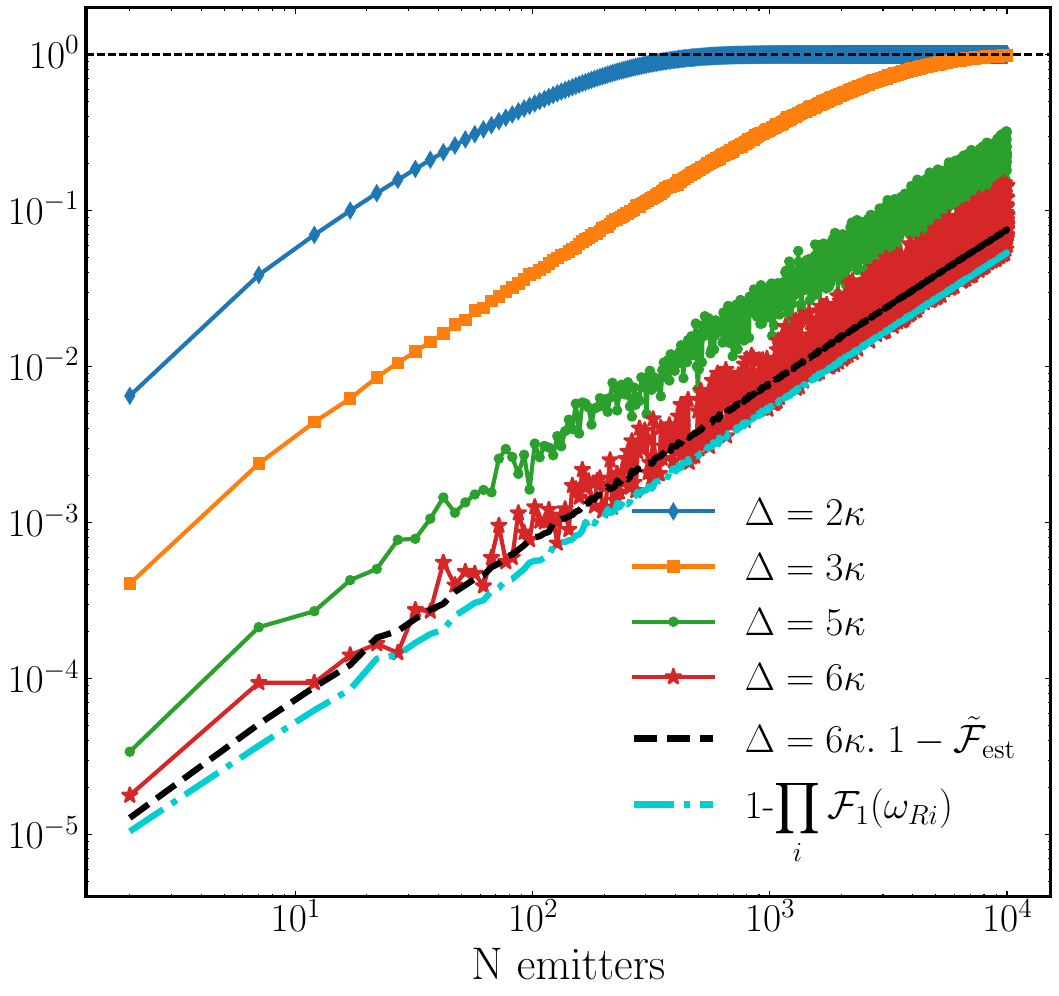}
\caption{
Heuristic infidelity scaling vs. number of transmitted qubits. We plot the estimates from Eq.~\eqref{eq:Nemitt_fid}, for multiplexed state transfer with frequency separations $\Delta=2\kappa, 3\kappa, 5\kappa$ and $6\kappa$. In the last case, we also plot the estimator~\eqref{eq:Nemitt_fid_est} based on photon wavepacket overlap (dashed line). Note how, for sufficient frequency separation, the infidelity approaches the limit imposed by single-qubit state transfer (dot-dashed line), which scales as $\mathcal{F}_1^N$.}
\label{fig4}
\end{figure}

Let us estimate the state transfer fidelity for multiple qubits from the numerical experiment with two emitters. The deviation from the ideal single-photon state transfer is captured in a $G(\omega_i, \omega_j)=\mathcal{F}_2(\omega_i,\omega_j)/\mathcal{F}_1(\omega_i)\mathcal{F}_1(\omega_j)$. In a pessimistic upper bound, one may assume that this correction factor describes the cross-talk between all pairs of quantum channels, introducing $N(N-1)/2$ corrections to the single-photon state transfer of $N$ qubits
\begin{equation}
  \label{eq:Nemitt_fid}
  \mathcal{F}_\text{est}(\omega_1,...,\omega_N)
   = \prod_i \mathcal{F}_1(\omega_i) \times \prod_{i<j} G(\omega_i, \omega_j).
\end{equation}
To achieve an asymptotic estimate of the cumulative effect of these corrections, one may use the fact that the overlap between single-photon modes almost fully explains $G$
\begin{equation} \label{eq:Nemitt_fid_est}
    \tilde{\mathcal{F}}_\text{est}(\omega_1,...,\omega_N) = \prod_i \mathcal{F}_1(\omega_i) \times \prod_{i<j} (1 - \left|\braket{\xi(\omega_i) |\xi(\omega_j)}\right|^2).
\end{equation}
Note that, despite the presence of $\mathcal{O}(N^2)$ factors, the overlap between two photons decreases exponentially with their separation in frequency space $|\omega_{Ri}-\omega_{Rj}|/\kappa$, a fact that allows further simplifications.

Both the upper bound $\mathcal{F}_\text{est}$ and the asymptotic formula $\tilde{\mathcal{F}}_\text{est}$ can be estimated using the simulation parameters from Sect.~\ref{sec:2phot}. For simplicity, our study assumes an equal spacing among the emitters in frequency space $\omega_{Ri+1}-\omega_{Ri}=\Delta$, with all frequencies centered around the original resonance. Exhaustive numerical simulations of the two-photon state transfer are used to compute the prefactor $G(\omega_i,\omega_j)\sim G(\omega_i-\omega_j) = G((j-i)\times \Delta)$ for all possible separations of $N$ qubits.

In Fig.~\ref{fig4} we compare the infidelity estimates for different mode spacings, $\Delta/\kappa=2,3,5$ and 6. The estimates based on exact simulations~\eqref{eq:Nemitt_fid} exhibit some oscillations due to the placement of filters in frequency space, but they all lay close to the exponential estimates coming from the overlap~\eqref{eq:Nemitt_fid_est}. In all cases, upper bounds and overlap estimates, the fidelity deteriorates exponentially with the growing number of emitters. However, the slope of the curve is proportional to $N$ and not $N^2$, indicating that out of the $N(N-1)/2$ correcting factor, only the nearest-neighbor overlap plays a significant role $\tilde{\mathcal{F}}_\text{est} \sim \mathcal{F}_1^N \times (1-\mathcal{I}_\text{overlap}(\Delta))^{N-1}$. Furthermore, as the spacing increases, the curves approach the limiting curve of $N$ independent state-transfer processes, identified by the dash-dot line in Fig.~\ref{fig4}. Indeed, for the spacing $\Delta=6\kappa$ discussed before, the infidelity is extremely close to the ideal case.

Out of these plots, we can extract the number of qubits that can be transmitted with a given error $\varepsilon$. However, given the simple behavior of $\tilde{\mathcal{F}}_\text{est}$ described above, this number may be estimated analytically from the mode overlap
\begin{equation}
N_\text{max} = \frac{\log(1-\varepsilon)}{\log[\mathcal{F}_1(1-\mathcal{I}_\text{overlap}(\Delta))]},
\end{equation}
with the corresponding bandwidth $\Delta N_\text{max}$. As an example, in $l_\mathrm{WG}=15$ m waveguide of Fig.~\ref{fig4}, a separation $\Delta > 6\times \kappa$ is very close to the single-photon limit. Under these conditions, and without other limitations, we could ideally transmit 500 qubits with an error below $10^{-3}$, using a bandwidth of $2\pi\times 5$ GHz. In practice, these numbers are unrealistic, but they highlight the potential of quantum links as common buses for transmitting information.

\subsection{Transfer optimization}

\begin{figure}
    \centering
\includegraphics[width=.9\columnwidth]{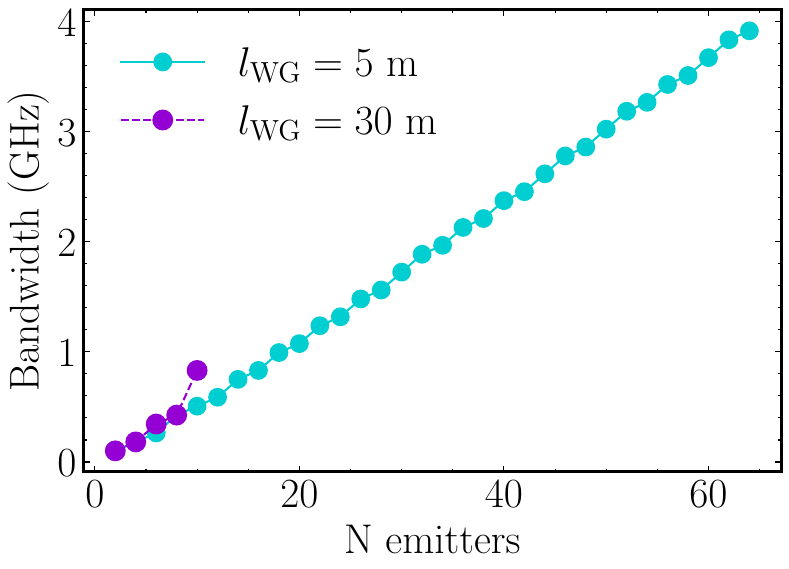}
\caption{Bandwidth requirements to host a particular number of emitters with a given global tolerance of $10^{-4}$ for the N-photons state transfer operation. The graph shows data for two different waveguides of $l_\mathrm{WG}=5$ m and $l_\mathrm{WG}=30$m. The lines stop when increasing the bandwidth would not improve the fidelity, that is, when the value of the single independent processes is itself greater than the tolerance imposed for the global gate, $1-\mathcal{F}(\omega_1)^N > \text{Tol.}$ These data have been obtained with the estimator $\tilde{\mathcal{F}}_\text{est}$ defined by equation~\eqref{eq:Nemitt_fid_est}, Unlike that of Fig.\ref{fig4} which came from two-photon simulations. }
\label{fig5}
\end{figure}

The multiplexed state transfer efficiency is ultimately limited by the fidelity of transmission of individual qubits $\mathcal{F}_1(\omega)$. Since this process has been extensively studied in Refs.~\cite{Penas2022} and \cite{Penas2023}, it makes sense to revisit these works, using those studies to optimize the single-photon state transfer and understand the actual information capacity of realistic waveguides.

Two distinct mechanisms dominate the single qubit state transfer infidelity: wavepacket distortion by propagation with non-linear dispersion relations and Stark shifts due to the time-varying controls $g(t)$. The wavepacket distortion may be corrected by imprinting phases in the emission and absorption controls $g(t)$~\cite{Penas2023}. Both effects can also be minimized by reducing the photon bandwidth $\kappa$ and enlarging the duration of state transfer---within the limits set by the emitters' intrinsic decoherence and dephasing.

Another limiting factor in multiplexed state transfer is the positioning of the emitters around the waveguide's resonances. This effect is particularly relevant for narrow photons and short waveguides, where the qubits can resolve the free spectral range. This effect accounts for the oscillations in the infidelity of multiplexed state transfer, both in Fig.~\ref{fig3} with two qubits and in the estimates for tens and hundreds of qubits in Fig.~\ref{fig4}. Optimally placing the emitters can substantially improve the transfer quality---see Fig.~\ref{fig3}, diamond dot. However, the optimal positioning of qubits limits the number of qubits transmitted in a waveguide within a given bandwidth.

Figure~\ref{fig5} illustrates the physical requirements for a setup in which the three optimizations---predistortion, wavepacket bandwidth, and emitter positioning---have been applied. The plot shows the number of qubits transmitted in waveguides of 15 and 30 meters, with a maximum error $\varepsilon = 10^{-4}$ estimated from  $\mathcal{F}_\text{est}$---which, as seen above, captures the overall scaling of resources very well. In both waveguides, the ultimate capacity is set by the single-qubit optimized infidelities $\mathcal{I}^1_{5 \text{m}} = 5\cdot 10^{-7}$ and $\mathcal{I}^1_{30 \text{m}} =10^{-5}$, as computed in Ref.~\cite{Penas2022}, with predistortion and optimal qubit positioning. These quantities already set a very stringent upper bond on the sequential and multiplexed transmission capacities. For instance, the 30 m waveguide only supports transmitting 10 qubits before the error exceeds the tolerance $10^{-4}$. Furthermore, while shorter waveguides allow high-quality transmission of more qubits---e.g., over 60 qubits for the 5 m waveguide---the bandwidth requirement of over 4 GHz imposes challenging design restrictions in the emitters and filter designs. Under these conditions, it may be advantageous, from an engineering perspective, to use multiple channels.

\section{Conclusions}\label{sec:conclusions}
This work has addressed the multiplexation of quantum information in a microwave quantum link that connects $N+N$ qubits in two quantum nodes, following setups from recent experiments~\cite{Kurpiers2017, Magnard2020, Storz2023}, but with tools that generalize to other waveguide-QED architectures.

The first important result is the demonstration that qubits can be transferred in different orthogonal modes by a suitable control of the interaction between the storage qubits and the quantum link. This required generalizing the methods for wavepacket engineering envisioned by~\cite{Cirac1995}, and extended and implemented in various works~\cite{Eichler2012, Pechal2013, Zeytinoglu2015, Kurpiers2017, Bienfait2019, Magnard2020, Storz2023, Qiu2023, Grebel2024}, arriving at close expressions for the controls to generate almost arbitrary wavepackets---under natural limitations of bandwidths set by the mediating cavities. As a particular case, the study demonstrates the accuracy of mode multiplexing for a family of sech-like pulses in a realistic setup.

The second important result is a study of frequency multiplexing for arbitrary numbers of qubits. An accurate analysis of the simultaneous transfer of two qubits identified cross-talk as the limiting factor in the process fidelity. Accurate simulations with two qubits in generic waveguides enable extrapolating the fidelity of multiplexed state transfer to higher numbers of qubits, showing that the single-qubit state transfer fidelity and the overlap between neighboring photon wavepackets ultimately limit this. Using heuristic scaling in combination with the physical constraints of state-of-the-art superconducting circuit experiments, we estimated the practical capacity of microwave waveguides for a given error tolerance. For instance, we found that as many as 60 photons could be sent through a 5-meter waveguide with a global infidelity below the usual threshold for fault-tolerant computation $10^{-4}$.

The tools developed in this work can be used to extend the capacity of waveguide-QED quantum links, guiding the design of superconducting connections between quantum computers, enabling the distribution of higher-dimensional quantum states~~\cite{Zhong2021, Qiu2023, Storz2023}, and extending the accessible Hilbert space in applications such as the generation of photonic cluster states~\cite{Schwartz2016, Ferreira2022}

\section{Acknowledgements}
This work has been supported by the European Union's Horizon 2020 FET-Open project SuperQuLAN (899354), the Spanish project PID2021-127968NB-I00 (MCIU/AEI/FEDER,EU), and Proyecto Sin\'ergico CAM 2020 Y2020/TCS-6545 (NanoQuCo-CM).

\appendix

\section{Orthogonal wavepackets}\label{app:orthogonal_gs}

As commented in the main text, we begin with one of the most commonly used wavepacket shapes in the literature, namely, the sech-like photon given by Eq.~\eqref{eq:sech-pulse}. The next mode is the antisymmetric wavefunction~\eqref{eq:first_ortho_photon} obtained by multiplying the sech-profile by a first-degree monomial $t^1$. In this case, the orthogonality is guaranteed by the difference in symmetry between both functions, $\int \xi_1(t)^*\xi_0(t)\mathrm{d}t=0$. For the second orthogonal photon we could naively increase the order of the monomial
\begin{align}
\xi_{\perp_2}(t)=\sqrt{\frac{15\kappa^5}{28\pi^4}}\sech(\kappa t/2) t^2 e^{-i\omega_Rt}.
\end{align}
However, this function is not orthogonal to $\xi_{0}$.  To ensure that the new proposed photon is orthogonal to all the others one can use the Gram-Schmidt method,
\begin{align}
    \xi_2(t)\propto\xi_{\perp_2}-\langle \xi_{\perp_2},\xi_0\rangle \; \xi_0-\langle \xi_{\perp_2},\xi_1\rangle \; \xi_1,
\end{align}
finally finding $ \xi_2(t) $ to be
\begin{align}
        \xi_2(t) = \frac{\sqrt{5\kappa}}{24 \pi^2 \kappa^2} \; \sech(\kappa t/2) \left(t^2-\frac{\pi^2}{3\kappa^2}\right)e^{-i\omega_Rt}.
\end{align}
Naturally, this technique can be extended to consider polynomials corrections of arbitrary order.

Any of these photon shapes can be plugged into \eqref{eq:gtmod} and \eqref{eq:phasedot}, and obtain from there the analytical formulas of the controls to produce them. In all cases the wavepacket is real $\xi_n\in\mathbb{R}$, and we can set $f_n=\xi_n,\;\theta_n=0$. However, as we remarked in the main text, the changes in sign in the envelope immediately leads us to obtain complex controls $g_n(t)$. The expressions for the controls of the first two orthogonal modes is given by
\begin{align}\label{eq:xi0_ap}
  g_0(t)&=\frac{\kappa}{2}\sech(\kappa t/2),~\mbox{and}\\
  \label{eq:g1_ap}
 g_1(t)&=\frac{\kappa(1+e^{\kappa t}+\kappa t)\sech(\kappa t/2) }{D_1},~\mbox{with}\\
 D_1 &= (1+e^{\kappa t})(-8{\rm Li}_2(-e^{-\kappa t}) +\kappa t (2\kappa t+8\log(1+e^{-\kappa t})+\nonumber \\
 & - \kappa t \sech^2(\kappa t/2)(1+\sinh(\kappa t) )))^{1/2},\nonumber
\end{align}
with the special function ${\rm Li}_n(x)=\sum_{k=1}^\infty x^k/k^n$.

\section{Scattering of orthogonal wavepackets}\label{app:scattering}

\begin{figure}
    \centering
    \includegraphics[width=\columnwidth]{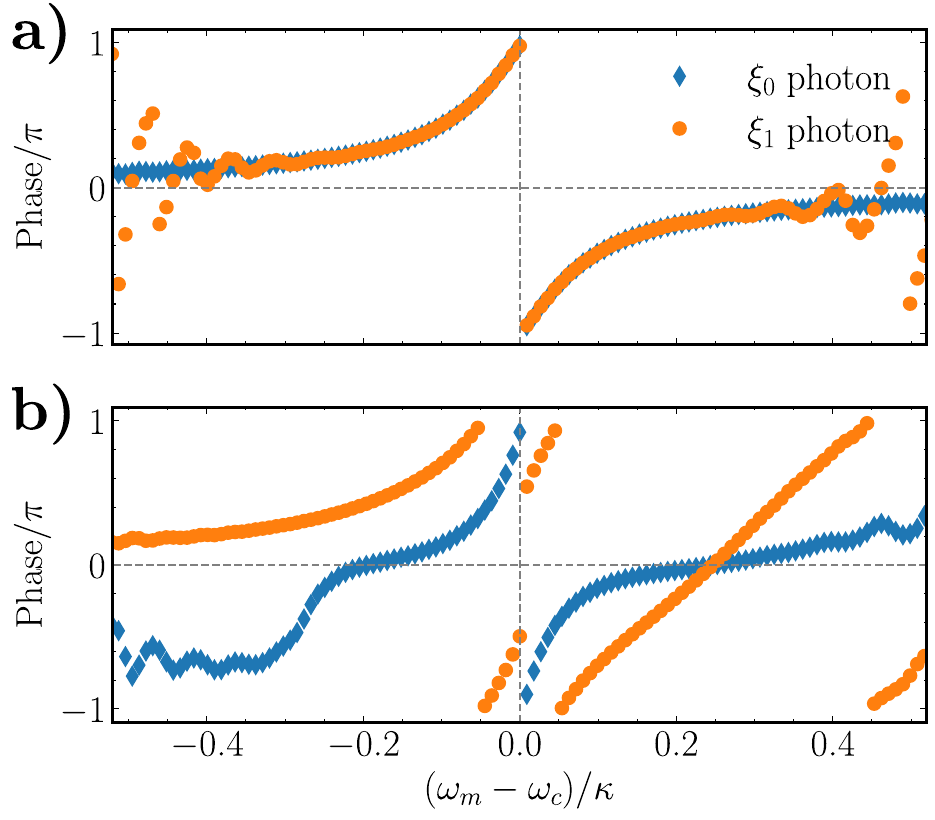}
    \caption{Panel a) shows the result of comparing the $\xi_0$ and $\xi_1$ photons after scattering off a resonator with no active control.
    Panel b) shows the phases acquired by $\xi_0$ after scattering off a qubit-resonator system with dynamic coupling $g_1(t)$ (blue dots). And that acquired by $\xi_1$ when at the other end $g_0(t)$ is applied (orange dots). Parameters of the simulation: $l = 30$ m, $\kappa / 2\pi = 60$ MHz. $N_{WG}= 300$ modes around a central frequency of $\omega_c/2\pi = 8.9$ GHz. Of those 300 modes only the 40 modes closest to the emitter's frequency are displayed, which correspond to a bandwidth of $\approx \kappa$.}
    \label{fig:scattering_ortho}
\end{figure}

The aim of this section is to examine what happens when a photon $\xi$ scatters off a qubit-resonator node which is using a control $g(t)$ designed to absorb a different photon mode. The analysis is based on numerical experiments with the two photon wavepackets $\xi_{0,1}(t)$ from the mode multiplexing section~\eqref{eq:sech-pulse}-\eqref{eq:first_ortho_photon}. The study compares the scattering phases acquired by the two modes in frequency space, first in an experiment with a bare resonator, and next with a resonator that interacts with a qubit using a time-dependent control $g(t)$. The numerical experiment is designed with a long enough waveguide, so that the photon fully fits in it and we can analyze the scattering state at a finite time $T$ after the collision, with the scattering state prior to the interaction with the second node. The comparison is performed in frequency space, where the probability of the different waveguide modes is undisturbed and only scattering phases are acquired.

Figure \ref{fig:scattering_ortho} illustrates the scattering phases after the interaction with a bare cavity (Panel a, top) and with a cavity and a qubit (Panel b, bottom). The phases acquired by the interaction with a resonator are independent of the incoming mode, $\xi_0(t)$ or $\xi_1(t)$, and are compatible with the expected phases from an input-output theory calculation~\cite{Penas2022}
\begin{equation} \label{eq:phase_scatt}
e^{i\phi_{\rm scatt}(\omega)}=\frac{i(\omega-\Omega_{\rm R2})+\kappa_2/2}{i(\omega-\Omega_{\rm R2})-\kappa_2/2}.
\end{equation}
Panel b) shows a completely different story. It displays the phase acquired by a photon in mode $\xi_0(t)$ when the cavity-filter interaction follows protocol $g_1(t)$ (blue dots) and the phase acquired by a photon in mode $\xi_1(t)$ when $g_0(t)$ is used (orange dots). Now the scattering phases are affected by the type of qubit-cavity interaction control, imprinting a distortion that is markedly different from the photon-cavity collisional phase. This said, the phase profiles are still computable and can be computed and used to learn the shape of the reflected photon, thus engineering other controls for later interactions.

\section{Mutual influence of the emitters.}\label{app:mutual_influence}

We observed from the simulations that the inclusion of the second node had consequences on the properties of the first and vice versa. This influence could be summarize as a recalibration of the frequency of the resonators, a modification of the decay times and a coherent exchange of population between them. Because all these effects occur only at the resonator-waveguide level, we chose to study them with an effective model ignoring the qubits. Also, for the sake of simplicity, we will drop the subindex R of the frequencies of the resonators throughout the appendix, and denote the waveguide frequencies as $\Omega_m$.  The Hamiltonian of such effective model is
\begin{align}\label{eq:bosonic_Ham}
   H = \sum_{i=1,2}\omega_{i} a^\dagger_i a_i +\sum_m \Omega_m b_m^\dagger b_m + \sum_{m,i=1,2} G_{m,i}\left(a^\dagger_i b_m +{\rm H.c.} \right).
\end{align}
We will break the calculation in two parts, first, we will depart from \eqref{eq:bosonic_Ham} and get rid of the waveguide modes, ideally obtaining an effective Hamiltonian for the resonators and a closed expression for the Lamb shift. Once we have such Hamiltonian, we would like to study how the $a_i^{\dagger} a_j$ with $i\neq j$ affects perturbatively to the energy of the resonators.

The first thing we do is to move to an interaction picture that rotates with the free energies of resonator and waveguide modes, that way, \eqref{eq:bosonic_Ham} becomes
\begin{align}
H^{I} = \sum_{m,i} G_{m,i} \left(a^\dagger_i b_m e^{+it(\omega_i-\Omega_m)} +a_i b^\dagger_m e^{-it(\omega_i-\Omega_m)} \right).
\end{align}
In this picture the equations of motion for the operators are
\begin{align}
\nonumber
    \dot{a}_j &= i \left[H,a_j\right]= -i \sum_m  G_{mj} b_m e^{+it(\omega_j-\Omega_m)},\\\nonumber
    \dot{b}_m &= i \left[H,b_m\right]= -i \sum_j G_{mj} a_j e^{-it(\omega_j-\Omega_m)} \longrightarrow\\\nonumber
    \longrightarrow b_m(t) &= b_m(0) -i \sum_j G_{mj} \int_0^t d \tau a_j(\tau) e^{-i\tau(\omega_j-\Omega_m)},
\end{align}
where we have changed the dummy index $i$ by $j$ for the sake of clarity.
We can substitute this formal solution for $b$ in the Hamiltonian
\begin{align}
H^{I} = -i \sum_{m,i,j}  G_{m,i}G_{mj} a^\dagger_i \int_0^t d \tau a_j(\tau) e^{-i\tau \omega_j}e^{+it\omega_i}   e^{-i\Omega_m(t-\tau)} \nonumber \\ +i \sum_{m,i,j}  G_{m,i}G_{mj} a_i   \int_0^t d \tau a^{\dagger}_j(\tau)  e^{+i\tau \omega_j}e^{-it\omega_i}   e^{+i\Omega_m(t-\tau)}.
\end{align}
Focusing now only in the first term
\begin{align}
H^{I} = -i \sum_{i,j} a^\dagger_i e^{+it\omega_i} \int_0^t d \tau  K(t-\tau) a_j(\tau) e^{-i\tau \omega_j} + \text{H.c.}\; ; \\   \quad \text{where} \quad K(t-\tau) = \sum_ mG_{m,i}G_{m,j} e^{-i\Omega_m(t-\tau)}.
\end{align}
Analogously to what is done in appendix B of \cite{ripoll2022quantum} (equation B.26). We can separate the time evolution of the operator $a$ as a slow contribution  $a_{j}(t)=a_{j,{\rm slow}}(t)e^{-i\omega_j' t}=a_{j,{\rm slow}}e^{-i\omega_j t}e^{-i\omega_j't}$ with an unknown frequency $\omega_j'$; doing this, and introducing $u = t - \tau$,
\begin{align}
H^{I} =  -i \sum_{i,j} a^\dagger_i e^{+it\omega_i} a_{j,{\rm slow}} e^{-i t \omega_j} \int_0^{+\infty} du  K(u) e^{i \omega'_j u}e^{-i \omega'_j t} +\text{H.c.}
\end{align}
Now, undoing the transformation to the interaction picture and using that $a_{j,{\rm slow}}e^{-i\omega_j' t}=a_j$,
\begin{align}
H^{I} = \sum_i \omega_i a^\dagger_i a_i  -i \sum_{i,j} a^\dagger_i a_j \int_0^{+\infty} du  K(u) e^{i \omega'_j u} +\text{H.c.},
\end{align}
which, by grouping terms appropriately leads to
\begin{equation}\label{eq:effective_H_exchange_terms}
            H_\text{eff} = \sum_i (\omega_i + \delta \omega_i)a^{\dagger}_i a_i  +  \tilde{G}\left( a_1 a_2^\dagger  + a_1^\dagger a_2\right),
\end{equation}
which is equation \eqref{eq:H_exchange_terms} of the main text once we recover the subindices R that we dropped, i.e., $ (\omega_i + \delta \omega_i) \longrightarrow (\omega_{Ri} + \delta \omega_{Ri}) $. The Lamb shift and effective coupling are defined through
\begin{align}
\delta \omega_i = \frac{1}{2 \pi} \int_0^{\infty} \int J_i^\text{QO}(\omega) e^{i(\omega'_i-\omega)u} d \omega du,
\end{align}
\begin{equation}
\tilde{G} = \frac{1}{2 \pi} \int_0^{\infty} \int \sqrt{J_i^\text{QO}(\omega)}\sqrt{J_j^\text{QO}(\omega)} \; e^{i(\omega'_i-\omega)u} d \omega du,
\end{equation}
\begin{align}
    \text{with} \quad  J_i^\text{QO}(\omega) = 2 \pi \sum_m |G_{m,i}|^2\delta(\omega-\Omega_m).
\end{align}
The Hamiltonian \eqref{eq:effective_H_exchange_terms} captures the physics of the renormalization of the frequencies due to the interaction of the resonator with the bath and an effective coherent exchange interaction between the resonators $i$ and $j$, also mediated by the waveguide. \eqref{eq:effective_H_exchange_terms} finishes the first part of the calculation and qualitatively explains the physics of the system in the region where the detuning is very small.

And now the second step begins, we are now interested in seeing what happens when the detuning is large and we would expect the exchange terms to become less relevant. We move once again to an interaction picture, now rotating at the dressed frequencies of the resonators $\omega^D_i = \omega_i+\delta \omega_i$, doing so, \eqref{eq:effective_H_exchange_terms} becomes
\begin{align}\label{eq:H_eff_eff}
            H_\text{eff} = \sum_i \omega^D_ia^{\dagger}_i a_i  +  \tilde{G}\left( a_1 a_2^\dagger  + a_1^\dagger a_2\right) \longrightarrow  \nonumber\\
            H^I_\text{eff} = \tilde{G}\left( a_1 a_2^\dagger e^{+it(\omega^D_2-\omega^D_1)}  + a_1^\dagger a_2 e^{-it(\omega^D_2-\omega^D_1)}\right).
\end{align}
Now, considering $\omega^D_2, \omega^D_1$ are sufficiently far apart we can perform a Magnus expansion up to second order to see how the effective coupling of the resonators $a_1, a_2$ may lead to a new renormalization of the frequencies, one on top of $\delta \omega_i$.


We can identify the second order term as
\begin{align}
    M_2(t) = \frac{1}{2} \int_0^t dt_1 \int_0^{t_1} dt_2 \left[H^I_\text{eff}(t_1), H^I_\text{eff}(t_2)\right],
\end{align}
which means computing the following ladder operators commutators
\begin{align}
\left[ a_1 a_2^{\dagger} e^{+i t_1 (\omega^D_2 - \omega^D_1)},  a_1^{\dagger} a_2 e^{-i t_2 (\omega^D_2 - \omega^D_1)} \right] \; \propto \; \left(- a_1^{\dagger} a_1 + a_2^{\dagger} a_2 \right), \\ \left[ a_1^{\dagger} a_2 e^{-i t_1 (\omega^D_2 - \omega^D_1)},  a_1 a_2^{\dagger} e^{+i t_2 (\omega^D_2 - \omega^D_1)} \right] \; \propto \; \left(+ a_1^{\dagger} a_1 - a_2^{\dagger} a_2 \right).
\end{align}
The integral coming from the first term is
\begin{align}
\frac{1}{2} \tilde{G} \left(- a_1^{\dagger} a_1 + a_2^{\dagger} a_2 \right) \int_0^t dt_1 \int_0^{t_1} dt_2  e^{+i t_1 (\omega^D_2 - \omega^D_1)} e^{-i t_2 (\omega^D_2 - \omega^D_1)} = \\ \frac{1}{2} \tilde{G} \left(- a_1^{\dagger} a_1 + a_2^{\dagger} a_2 \right) \int_0^t dt_1 \frac{i}{\omega^D_2-\omega^D_1} \left(1-e^{+i t_1 (\omega^D_2 - \omega^D_1)}\right) = \\
= -i \frac{1}{2} \frac{\tilde{G}^2}{\omega^D_2-\omega^D_1}\left( a_1^{\dagger} a_1 + (-a_2^{\dagger} a_2) \right) t + (\text{terms} \not\propto  t).
\end{align}
A minus sign from the exponential cancels out with the minus sign in the second commutator, giving the same result. Given this, the final Hamiltonian up to second order in the Magnus expansion has simply the form
\begin{align}\label{eq:eff_eff_Hamiltonian}
    \tilde{H}_\text{eff} = \sum_i C_i a_i^{\dagger} a_i, \; \text{with} \; C_i = (-1)^{i+1} \frac{\tilde{G}^2}{\omega^D_2-\omega^D_1},
\end{align}
which is equation \eqref{eq:eff_eff_Hamiltonian_main} of the main text once $\omega^D_2-\omega^D_1 \longrightarrow \omega^D_{R2}-\omega^D_{R1}$.
What \eqref{eq:eff_eff_Hamiltonian} tells us is that there is a frequency shift $C_i$ affecting the ith resonator due to its interaction with the other (or others), note that this shift is different from the Lamb shift, which is already taken into account in the definition of the dressed frequencies $\omega^D_i$. Furthermore, since $\tilde{G}$ is a constant representing the effective interaction strength between resonators, this mutually induced frequency shift is $C_i \propto \left( \omega^D_2-\omega^D_1 \right)^{-1} $. This behavior is observed in the simulations with high level of accuracy as it is shown in Fig.~\ref{fig:induced_change_kappa}.

\begin{figure}[t]
    \centering
\includegraphics[width=\columnwidth]{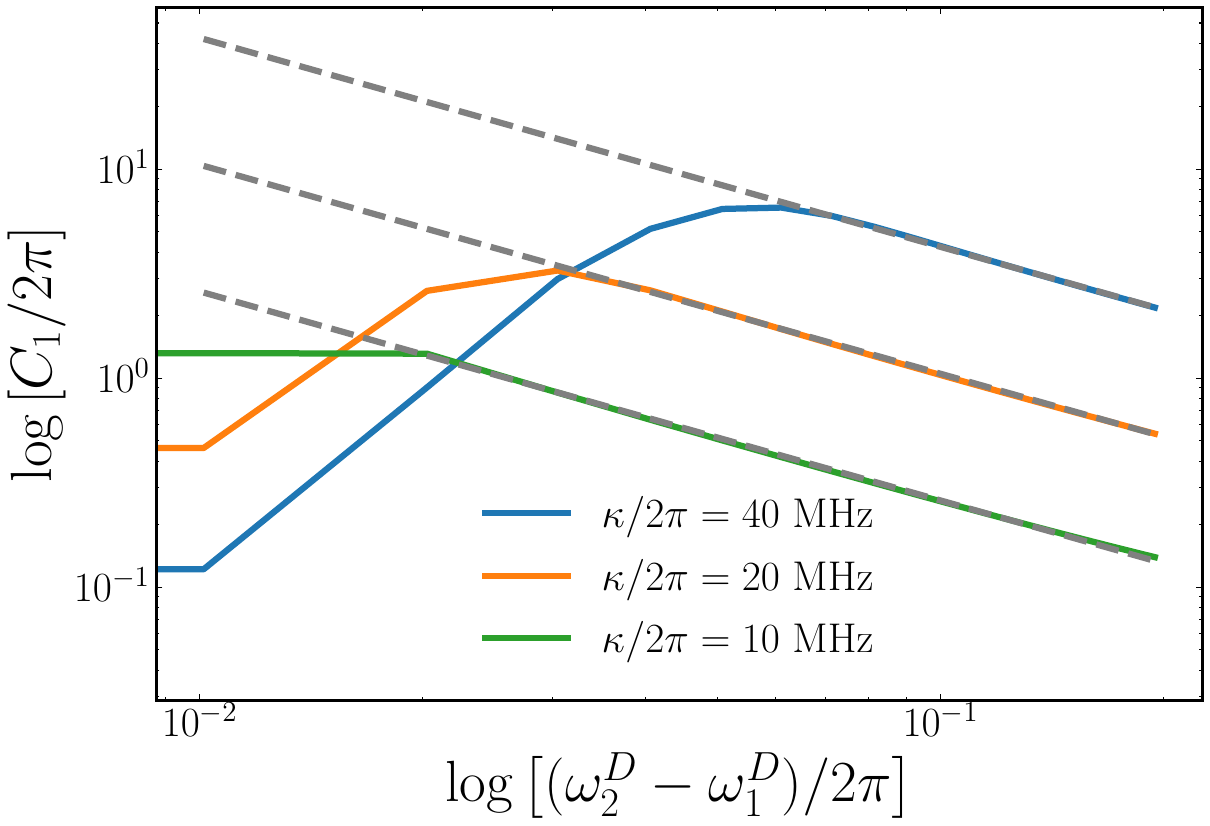}
\caption{Frequency shift of the resonator 1 (arbitrary labeling) induced by the presence of the physically identical resonator 2 for three different values of the $\kappa$. We can estimate the value of $\tilde{G}^2$ from the slope of the dashed lines. The parameters of the simulation are $l= 30$m to avoid discrimination of the mode structure of the waveguide by any of the three resonator widths. We do spectroscopy of the resonator with a depletion experiment using as our control $g(t) = \kappa_\text{eff}/2 \text{sech} (\kappa_\text{eff} t / 2)$, where $\kappa_\text{eff} = \kappa$ }
\label{fig:induced_frq_shift}
\end{figure}

\begin{figure}[b]
    \centering
\includegraphics[width=\columnwidth]{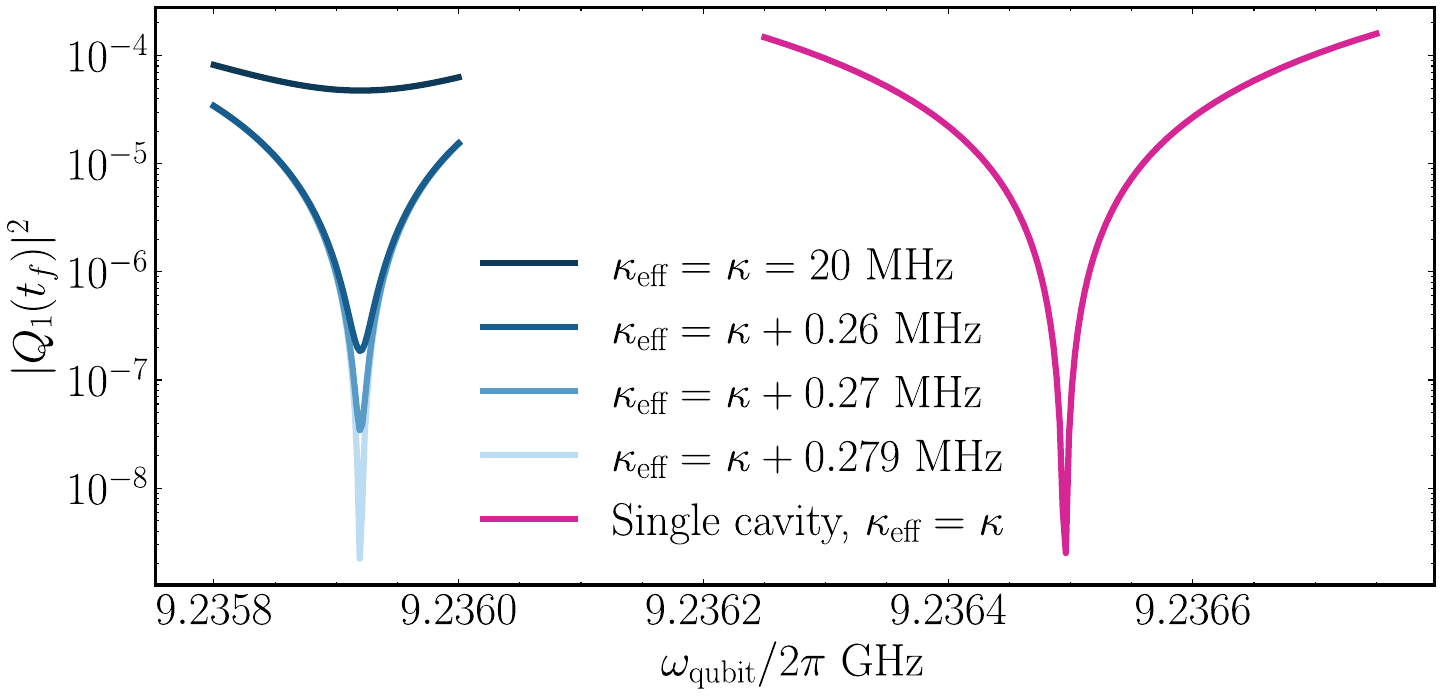}
\caption{Depletion of qubit 1 (arbitrary labeling) when there is only a resonator connected at node 1 (pink line) versus when there is another one (blue lines). In all cases with two resonators they are connected at the same spacial points but with a large detuning of $(\omega_1-\omega_2)/2\pi = 200$ MHz, or around $20\kappa$. We run a qubit depletion experiment using as our control $g(t) = \kappa_\text{eff}/2 \text{sech} (\kappa_\text{eff} t / 2)$. This is all computed after the $\delta \omega$ has already been optimized. $l=30$ m, $N_\text{modes} = 390$.}
\label{fig:induced_change_kappa}
\end{figure}

Fig.~\ref{fig:induced_frq_shift} shows how after an initial phase in which the physics is dominated by the exchange terms and a spectroscopy by means of depletion does not make much sense, when $(\omega_2^D - \omega_1^D) \approx 2\kappa $ the mutually induced frequency shift closely follows the prediction of \eqref{eq:H_eff_eff}, represented by the straight dashed lines. This is remarkable, since the effective model had been calculated only taking the resonators into account and the depletion experiment of Fig.~\ref{fig:induced_frq_shift} is done with the complete Hamiltonian of the problem. This is proof that it is a good approximation and it brings valuable insights.

Since both the values of $\delta \omega$ and $\kappa$ are ultimately related with the spectral function $J^\text{QO}$. It is expected that the later also changes by the mutual influence between resonators. Fig.~\ref{fig:induced_change_kappa} shows how this is indeed the case. Just as $\delta_i$ has to be carefully calibrated to much $\omega_i$ so $\kappa_\text{eff}$ has to be.  As we can see in Fig.~\ref{fig:induced_change_kappa}, a miscalibration of this parameter of just around $1 \%$ leads to 5 orders of magnitude in depletion efficiency.

\newpage

\bibliography{paper}

\end{document}